\documentclass[graybox]{svmult}

\usepackage{mathptmx}       
\usepackage{helvet}         
\usepackage{courier}        
\usepackage{type1cm}        
\usepackage{makeidx}         
\usepackage{graphicx}        
\usepackage{multicol}        
\usepackage[bottom]{footmisc}

\usepackage{amssymb,amsmath}

\makeindex             

\begin{document}

\title*{Testing Modified Gravity with Gravitational Wave Astronomy}
\author{Carlos F. Sopuerta and Nicol\'as Yunes}
\institute{Carlos F. Sopuerta \at Institut de Ci\`encies de l'Espai (CSIC-IEEC), 
Campus UAB, Torre C5 parells, 
Bellaterra, 08193 Barcelona (Spain), \email{sopuerta@ieec.uab.es}
\and 
Nicol\'as Yunes \at 
Princeton University, Physics Department, Princeton, NJ 08544 (USA), \email{nyunes@princeton.edu}}
%
%
\maketitle

\abstract*{The emergent area of gravitational wave astronomy promises to
provide revolutionary discoveries in the areas of astrophysics, 
cosmology, and fundamental physics.  One of the most exciting possibilities
is to use gravitational-wave observations to test alternative theories
of gravity.  In this contribution we describe how to use observations
of extreme-mass-ratio inspirals by the future Laser Interferometer Space Antenna
to test a particular class of theories:  Chern-Simons modified gravity.}

\abstract{The emergent area of gravitational wave astronomy promises to
provide revolutionary discoveries in the areas of astrophysics, 
cosmology, and fundamental physics.  One of the most exciting possibilities
is to use gravitational-wave observations to test alternative theories
of gravity.  In this contribution we describe how to use observations
of extreme-mass-ratio inspirals by the future Laser Interferometer Space Antenna
to test a particular class of theories:  Chern-Simons modified gravity.}

\section{Introduction} \label{introduction}

Gravity, as we see it from our four-dimensional spacetime perspective, appears
as the weakest of all physical interactions known to date.    Despite this fact,
it is the force that governs the large-scale structure of the universe.  In the
context of Einstein's theory of relativity, gravity also determines the spacetime
geometry and hence the relations between the events that take place on it.

As is well known, Newtonian mechanics together with Newton's law of gravitation are sufficient
to describe a wide range of phenomena governed by gravity, from the motion of objects
near the surface of our planet Earth to the motion in the Solar system, and even 
at much larger scales.  Relativistic effects show up when we make very precise observations
of astronomical systems, and also of systems that involve either strong gravitational fields 
or fast motions.  However, due to the weakness of gravity, these effects are
difficult to measure with present technology and, as a consequence, only certain
regimes of relativistic gravity have been tested so far.  These tests confirm, to their level 
of precision, the validity of general relativity (GR) (see~\cite{lrr-2006-3} for a review).  
These experimental tests include observations of the motion of different objects in the solar system and observations 
of millisecond binary pulsars.   In the first case, a dimensionless measure
of the Newtonian gravitational potential yields\footnote{Here, $G$ denotes the gravitational
Newton constant, $c$ the speed of light, $M^{}_\odot$ the mass of the Sun, and $r$ different
distance measures.}
\begin{eqnarray}
\frac{\Phi^{}_{Newtonian}}{c^2} = \frac{G M^{}_\odot}{c^2\; 1 {\rm AU} } \sim 10^{-8}\,.
\end{eqnarray}
In the second case~\footnote{We here choose data from the well-known Hulse and Taylor
binary pulsar (PRS B1913+16)~\cite{Weisberg:2004hi}, the one that provided the first 
strong evidence for the existence of gravitational waves (GWs)~\cite{Hulse:1974eb}.},
the same dimensionless measure yields
\begin{eqnarray}
\frac{\Phi^{}_{Newtonian}}{c^2} \sim \frac{G M^{}_\odot}{c^2\; 
r^{\rm periastron}_{\rm Hulse-Taylor} } \sim 10^{-6} \,.
\end{eqnarray}
We can compare these numbers with an estimation for the case of binary black holes (BHs)
near the merger phase
\begin{eqnarray}
\frac{\Phi^{}_{Newtonian}}{c^2} \sim \frac{G M^{}_{\rm BH}}{c^2\;(\mbox{a few } 
r^{}_{\rm Horizon}) } \sim 10^{-1} - 1 \,. \label{extremegravity}
\end{eqnarray}
This indicates that despite the accurate measurements achieved up to now,
both from the solar system and from pulsars,
there is still a long way to go until we can have observations of situations 
with truly strong gravitational fields.  This means that there are sectors of 
the gravitational theory that we have not yet tested.  In other words, we 
know that General Relativity correctly describes Nature in certain
regimes, but do not know whether this is also the case when gravitational fields are extreme, 
in the sense of Eq.~(\ref{extremegravity}), where alternative theories of gravity might be relevant.  

To access the gravitational regime not yet tested one can try to resort to
electromagnetic observations (see~\cite{lrr-2008-9} for a review) as new
observatories in the high-energy end of the spectrum have good potential
for such a goal.  Another possibility is to resort to a different messenger,
namely GWs, or a combination of electromagnetic and GW observations through
multi-messenger astronomy in the future.  Gravitational Wave Astronomy (GWA) 
is an emergent area that promises to bring revolutionary discoveries to the areas of astrophysics,
cosmology, and fundamental physics.  There are a number of ground
laser interferometric detectors (LIGO~\cite{LIGOsimplified}, VIRGO~\cite{VIRGO}, etc.) that will detect GWs, in the high frequency range ($f \sim 10 - 10^{3}$ Hz) during the
next decade. There are also ongoing developments for future detectors in space, like the Laser Interferometer Space Antenna (LISA)~\cite{LISAsimplified} or DECIGO~\cite{DECIGO}.  
In particular, LISA will operate in the low frequency band ($f \sim 10^{-4} - 1$ Hz), a band not
accessible from the ground due to seismic noise, and probably the richest band in terms of
interesting astrophysical and cosmological sources.

Apart from these detectors, there is work in progress to use networks of
millisecond pulsars to detect GWs in the ultra-low frequency range
($f \sim 10^{-9} - 10^{-8}$ Hz).  Millisecond pulsars have already being
used to test alternative theories of gravity, like scalar-tensor
theories (see, e.g.~\cite{Damour:2007ti}).  These pulsars are remarkable stable rotators, 
and as such, they require only simple models to describe their spin-down and 
times-of-arrival (TOAs) with a precision of $< 1$ $\mu$s over many years of observations.
GWs are not included in the analysis, so their existence will 
induce differences between the measured and theoretical TOAs, the so-called
{\em timing residuals}. 
To determine the exact origin of the timing residuals, that is, effects different
from GWs (calibration effects, errors in planetary ephemeris, irregularities in
the pulsar spin-down, etc.) it is necessary to correlate the timing residuals of
multiple pulsars.  It has been estimated that with a timely progress in 
technology, a successful detection of GWs should happen within a
decade~\cite{Verbiest:2009av} (or alternatively the experiments will rule out current 
predictions for GW sources in this frequency band).

GWs are a double-edge tool: On the one hand, their detection is quite a difficult problem that
requires very advanced technology. On the other hand, they are an ideal tool to test 
strong gravity (in the sense of Eq.~\ref{extremegravity}),
since GWs carry almost uncorrupted information from their sources.   
In the next sections we discuss the following points: (i) The basics of the planned 
LISA mission; (ii) The main properties of EMRIs and their GW emission; (iii) How to
use EMRIs to test alternative theories of gravity with a focus on Dynamical Chern-Simons 
Modified Gravity (DCSMG). This discussion is  based on work described 
in~\cite{Sopuerta:2009iy}.

\section{LISA: The Laser Interferometer Space Antenna}

LISA~\cite{LISAsimplified} is a joint NASA-ESA mission designed to 
detect and analyze the gravitational radiation coming from astrophysical and 
cosmological sources in the low-frequency band (corresponding to oscillation periods 
in the range $10$ s -  $10$ hours).  LISA consists of three identical spacecrafts flying 
in a triangular constellation, with equal arms of $5\cdot 10^{6}$ km each. As GWs 
from distant sources reach LISA, they warp space-time (locally generating curvature), 
stretching and compressing the triangle.  Thus, by precisely monitoring the separation 
between the spacecrafts, we can measure the GWs, and their shape and timing teach us
about the nature and evolution of the systems that emitted them.

\begin{figure}[b]
\sidecaption
\includegraphics[scale=.45]{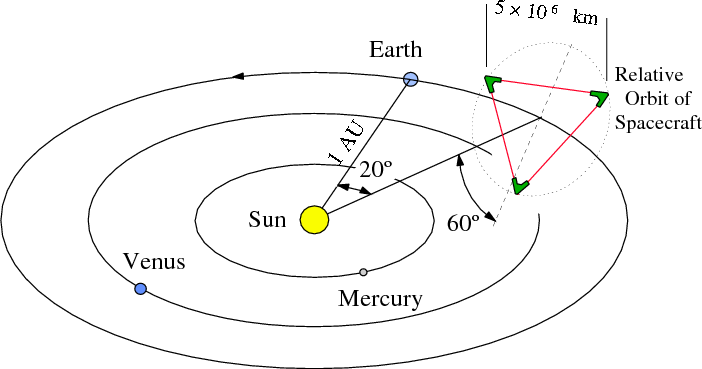}
\caption{Figure reproduced from~\cite{lrr-2000-3} with permission. It illustrates the configuration
of the motion of the LISA constellation.}
\label{fig:lisaorbit}       
\end{figure}

The LISA constelation is in orbit around the Sun, at a plane inclined 
by $60$ degrees to the ecliptic. The triangle appears to rotate once around its center 
in the course of a year's revolution around the Sun (see Fig.~\ref{fig:lisaorbit}). 
The center of the LISA triangle traces an Earth-like orbit in the ecliptic plane, trailing Earth
by $20$ degrees. The free-fall orbits of the three spacecraft around the Sun 
maintain this triangular formation, with the triangles appearing to rotate about its center 
once per year.

\begin{figure}[b]
\sidecaption
\includegraphics[scale=0.84]{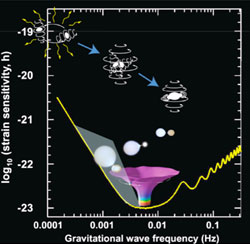}
\caption{This figure shows the LISA sensitivity response function in terms of the frequency.  
It allows shows the main sources of GW for LISA: (i) Massive BH mergers, entering the band from
the upper left corner (inspiral phase) and evolving in frequency until they merger and the
final BH rings down.  (ii) Galactic binaries. They are almost monochromatic sources and there
many of them, forming a foreground from which only a fraction can be individually distinguished.
(iii) EMRIs.  The capture and subsequent inspiral of a stellar-mass object into a Massive BH (see 
Sect.~\ref{emris}).}
\label{fig:lisasensitivity}       
\end{figure}

The sensitivity of LISA as a GW observatory is described by the strength of its response to impinging GWs as a function of frequency. At low frequencies it is limited
by acceleration noise; at mid frequencies by laser shot noise and optical-path measurement errors;
and at high frequencies by the fact that the GW wavelength becomes shorter 
than the LISA arm length, reducing the efficiency of the interferometric measurement 
(see Fig.~\ref{fig:lisasensitivity}).

Whereas ground-based detectors will make the first detections
in the high-frequency band and also inaugurate the field of GWA, space-based detectors 
like LISA, operating in the low-frequency band, will enable us to explore this new field in detail.
In this band, there are several important sources of GWs (see Fig.~\ref{fig:lisasensitivity}), 
such as massive BH mergers, i.e.,~mergers of BHs that grow in galactic centers 
and become a binary after their host galaxies collide. Another LISA source of GWs are 
millions of galactic binaries, which form a GW foreground
(except for the more bright ones, which can be separately identified as foreground).
Yet another important source is the capture and inspiral of stellar-mass compact
objects (white dwarfs, or neutron stars, or stellar BHs) by massive
BHs at galactic centers.  This source, commonly referred to as Extreme-Mass-Ratio 
Inspirals (EMRIs), is the one we will focus on in this contribution as a high-precission 
tool for fundamental physics tests.  Apart from these three sources, there are also prospects 
of detecting eventual stochastic GW backgrounds from the early universe (inflation,superstrings,topological defects, etc.).

\section{Extreme-Mass-Ratio Inspirals: EMRIs}
\label{emris}

EMRIs are composed of a stellar-mass compact object (SCO) spiraling into a massive BH 
(MBH) located in a galactic center.  Since the masses of interest for the SCO are around 
$m^{}_{\star} = 1-10^2\; M^{}_{\odot}$, 
and for the MBH are in the range $M^{}_{\bullet}= 10^5-10^7\; M_{\odot}$, the mass-ratio for these systems is  
$\mu=m^{}_{\star}/M^{}_{\bullet} \sim 10^{-7} -10^{-3}$.  During the inspiral phase, an EMRI losses energy and angular momentum 
via the emission of GWs, producing a shrinking of the orbit, which means that
the period decreases and, as a consequence, the GW frequency increases. 

There are several astrophysical mechanisms that have the potential to produce EMRIs (see~\cite{AmaroSeoane:2007aw} 
for a review on EMRI astrophysics and other aspects).
The most studied mechanism is based on the properties of the SCO's dynamics in stellar cups
around MBHs at galactic centers.  There is a small but non-negligible probability that one of
these objects may {\em fall} into a bounded trajectory with respect to the MBH, due to gravitational interactions with other bodies. If so, GW emission would then force the system to decay and 
plunge into the MBH in a period significantly smaller than the Hubble time scale. 
Initially, the orbit can be quite eccentric, with eccentricities in the range 
$1-e \sim 10^{-6}-10^{-3}$, but by the time they enter into the LISA band the eccentricity is expected to be substantially reduced (due to GW emission), although
it will probably still be significant (in the range $e \sim 0.5-0.9$).  A remarkable fact about EMRIs is that during the last year before plunge they emit on the order of $10^{5}$ or more
GW cycles, carrying a lot of information about the MBH strong field region.

Moreover, it has been estimated that LISA 
will be able to detect GW signals of around $10-10^3$ EMRIs per year up to distances of 
$z \lesssim 1$~\cite{Gair:2004ea,Hopman:2006xn}.  
These signals will be hidden in the LISA instrumental noise and in the GW foreground 
produced mainly by compact binaries in the LISA band.  Thus, in order 
to extract the EMRI signals we need a very accurate theoretical description of the gravitational 
waveforms.  The main difficulty in producing these is in the description of the 
gravitational effects of the SCO on its own trajectory.  These effects produce deviations
in the SCO's motion away from a geodesic around the MBH. One can think of such a deviation as induced by the action of a local force, the so-called \emph{self-force}. Analogously, one can think of the SCO as moving on a geodesic of the spacetime generated both by the MBH and the SCO itself.
There is an entire research program devoted to the computation of the self-force carried out by a community that annually gathers at the CAPRA Ranch Meetings on Radiation Reaction. 
In the last few years, there has been tremendous
progress in this direction, among other things leading to the first computations of
the gravitational self-force for generic orbits around a non-spinning MBH.   Details on the 
self-force research program and recent advances can be found in the 
reviews~\cite{Barack:2009ux,Poisson:2004lr} and references therein.

LISA observations of EMRIs have the potential to make revolutionary discoveries in Astrophysics,
Cosmology, and Fundamental Physics.  With regard to the first, Astrophysics, we expect the following discoveries: 
to better understand the dynamics in galactic centers (mass segregation, resonant relaxation, 
massive perturbers, etc); to obtain information about the mass spectrum of stellar BHs 
in galactic nuclei in order to understand the formation of stellar BHs and their relation to 
their progenitors; to obtain information on the distribution of the MBH spins for masses up to a 
few million solar masses, which has implications for galaxy formation models; perhaps to detect 
the inspiral of an Intermediate-Mass BH (IMBHs) into a MBH, which will give direct evidence for 
the existence of IMBHs; etc.  Regarding Cosmology, it has been proposed~\cite{MacLeod:2007jd} 
that precise measurements of the Hubble parameter are possible by correlating LISA EMRI observations (which act as standard ``sirens'' and provide precise measurements of luminosity distances up to $z\sim 1$) 
with galaxy redshift surveys, which would provide statistical redshift information of the EMRI events
(which cannot be inferred from the GW observations).  Applying this idea to a simplified cosmological model, it has been estimated that using $20$ or more EMRI events to $z \sim 0.5$, one could measure the Hubble constant to better than one percent precision.

Finally, and of most relevance for this contribution, EMRIs have also a great potential for Fundamental Physics, based on the fact that EMRI GW are long and carry a detailed 
{\em map} of the MBH spacetime, i.e.,~of the MBH multipole moments.
It has been estimated that LISA can measure the main parameters of an EMRI system with high 
precision~\cite{Barack:2003fp}:
\begin{equation}
\Delta(\ln M^{}_\bullet)\,,~\Delta\left(\ln\frac{m^{}_{\star}}{M^{}_\bullet}\right)\,,~
\Delta\left(\frac{S^{}_\bullet}{M^2_\bullet}\right)~\sim~10^{-4}\,,~~
\Delta\Omega~\sim~10^{-3}\,, 
\end{equation}
where $S^{}_\bullet$ denotes the MBH spin and $\Omega$ the solid angle (related to sky 
localization of the source).  It is also expected (see~\cite{Ryan:1997hg,Barack:2006pq}) 
that LISA will be able to measure $3-5$ MBH multipole moments with good accuracy.
Therefore, EMRI LISA observations provide a unique 
opportunity to test the no-hair theorem and also to perform tests of alternative theories of gravity,
which is the main subject of the remainder of this contribution.

\section{Testing Theories of Gravity with EMRIs: EMRIs in DCSMG}

There are many modifications of General Relativity available in the literature, and hence in principle
it seems difficult to justify a particular choice.  However, not all theories are created equal. In 
particular, not all theories available are consistent in all regimes; in many of them the status 
of BH solutions is unclear, in the sense that either the solutions are not known or, if they are known, 
it is unclear that they are unique or that they represent the final state of gravitational collapse. 
On the other hand, in order to gain some insight on the effects modifications of gravity may have on 
GW observations it is good to explore several different particular theories.  In this sense, 
one can also propose certain criteria for a theory to be a reasonable candidate to test GR 
with LISA~\cite{Yunes:2009ke}. 

An example of such a theory is DCSMG (see~\cite{Alexander:2009tp} for a recent review). This modification to General Relativity was introduced by Jackiw and Pi~\cite{jackiw:2003:cmo} and it
consists of the addition of a new term to the Einstein-Hilbert Lagrangian that generalizes the
standard 3-D Chern-Simons term. This new term is a parity violating interaction that is motivated by 
several quantum gravity approaches, like string theory and loop quantum gravity.  
This modification is also motivated from an effective field theory standpoint, through the inclusion
of high-curvature terms to the action (see~\cite{Weinberg:2008hq} for an application of this
approach to inflationary cosmology).

In this 4D theory, the action is given by:
\begin{equation}
{\cal S} = {\cal S}^{}_{\rm EH} + {\cal S}^{}_{\rm CS} +  {\cal S}^{}_{\phi} + {\cal S}^{}_{\rm matter}\,,
\end{equation}
where ${\cal S}^{}_{\rm EH}$ is the Einstein-Hilbert action
\begin{equation}
S^{}_{\rm{EH}} = \kappa \int d^4x  \; \sqrt{-\mbox{g}}  \; R\,,  \quad\quad \kappa = \frac{1}{16 \pi G}\,,
\end{equation}
which is modified by the addition of a term containing the Pontryagin density 
(${\,^\ast\!}R\,R = R^{}_{\alpha \beta \gamma \delta}{\,^\ast\!}R^{\alpha \beta \gamma \delta} =
\frac{1}{2} \epsilon^{\alpha \beta \mu \nu} R^{}_{\alpha \beta \gamma \delta}R^{\gamma \delta}{}^{}_{\mu \nu}$)
\begin{equation}
{\cal S}^{}_{\rm{CS}}= \frac{\alpha}{4} \int d^4x \; \sqrt{-\mbox{g}} \;\phi \; {\,^\ast\!}R\,R\,.
\label{actionCS}
\end{equation}
Here, the  Pontryagin density, a topological invariant in 4D, is multiplied by a scalar field, $\phi$,
to produce a modification of the GR field equations, which is proportional to the coupling constant $\alpha$.  
In addition we have the action term of the
scalar field:
\begin{equation}
{\cal S}^{}_{\phi} = - \beta \int d^{4}x \; \sqrt{-\mbox{g}} \; \left[ \frac{1}{2} \mbox{g}^{\mu \nu}
\left(\nabla_{\mu} \phi\right) \left(\nabla_{\nu} \phi\right) + V(\phi) \right] \,,
\label{actionphi}
\end{equation}
where $\beta$ is another coupling constant.
In the original version of the theory, the scalar field $\phi$ was forced to be a fixed function, devoid of dynamics and a contribution to the action. This leads to an additional constraint, the vanishing of the
Pontryagin density, which is too restrictive.  In particular, it disallows spinning BH solutions for scalar fields
whose gradient is time-like~\cite{Grumiller:2007rv} and forbids perturbations of non-spinning BHs~\cite{Yunes:2007ss}.  
Finally, ${\cal S}^{}_{\rm matter}$ is the action of any additional matter fields.

We now summarize the main results on the study of EMRIs in DCSMG~\cite{Sopuerta:2009iy}.  The first point
is that spinning MBHs are no longer described by the Kerr metric (although non-spinnning ones are described
by the Schwarzschild metric).  
Using the small-coupling and slow-rotation approximations, the exterior, stationary and axisymmetric
gravitational field of a rotating BH in dynamical DCSMG modified gravity, in Boyer-Lindquist type 
coordinates, is given by~\cite{Yunes:2009hc}:
\begin{equation}
ds^{2} = ds^{2}_{\rm Kerr} + \frac{5 \xi a}{4 r^{4}} \left[1 + \frac{12 M}{7 r} + \frac{27 M^{2}}{10 r^{2}} \right] 
\sin^{2}{\theta} dt d\varphi\,, \label{kerrcs}
\end{equation}
where $ds^{2}_{\rm Kerr}$ is the line element for the Kerr metric, $M$ and $a$ are the MBH mass and spin parameter, 
and $\xi= \alpha^{2}/(\kappa \beta)$ [see Eqs.~(\ref{actionCS}) and (\ref{actionphi})]. 
The multipolar structure of the modified metric remains completely
determined by only two moments (no-hair or two-hair theorem): the mass monopole and the current dipole. The relation, however, 
between these two moments and higher-order ones is modified from the GR expectation at multipole $\ell \geq 4$.
On the other hand, the solution for the DCSMG scalar field $\phi$ is:
\begin{equation}
\phi =  \frac{5}{8} \frac{\alpha}{\beta} \frac{a}{M} \frac{\cos(\theta)}{r^2} 
\left(1 + \frac{2 M}{r} + \frac{18 M^2}{5 r^2} \right)\,,
\end{equation}
which is axisymmetric and fully determined by the MBH geometry~\cite{Yunes:2009hc}.
Hence, the no-hair theorem still holds in this theory.

Regarding the equations of motion for the SCO, it has been shown~\cite{Sopuerta:2009iy} that point-particles follow 
geodesics in this theory, as in GR.  Moreover, it turns out that the metric given in Eq.~(\ref{kerrcs}) has the 
same symmetries as the Kerr metric (stationary and axisymmetric), including the existence of a 2-rank Killing
tensor.  As a consequence, the geodesics equations are also fully integrable, and the difference
with respect to Kerr can be encoded in a single function~\cite{Sopuerta:2009iy}.
One can see that the innermost-stable circular orbit (ISCO) location is DCSMG shifted by~\cite{Yunes:2009hc}:
\begin{equation}
R^{}_{\mbox{\tiny ISCO}}= \underbrace{6M \mp \frac{4\sqrt{6}a}{3}-\frac{7a^2}{18 M}}_{\mbox{GR Piece}}
         \pm \underbrace{\frac{77\sqrt{6}a}{5184} \frac{\alpha^{2}}{\beta \kappa M^{4}}}_{\mbox{CS Modification}}
\end{equation}
where the upper (lower) signs correspond to co- and counter-rotating geodesics. Notice that the 
DCSMG correction acts {\emph{against}} the spin effects.  One can also check that the three fundamental
frequencies of motion~\cite{Schmidt:2002qk} change with respect to the GR values.

The next important question to address is how GW emission and propagation is affected in DCSMG.  First of all, it has been shown~\cite{Sopuerta:2009iy} that observers far away from the sources can only observe
the same polarizations as in GR, although there is an additional mode, a breathing mode, that has an impact
in the strong-field dynamics but decays too fast with distance to be observable in the GW emission.
The DCSMG EMRI analysis of~\cite{Sopuerta:2009iy} was carried out in the so-called
\emph{semi-relativistic} approximation~\cite{Ruffini:1981rs}, where the motion is assumed geodesic 
and GWs are assumed to propagate in a flat spacetime.  
Neglecting \emph{radiation reaction} effects,
the dephasing between DCSMG and GR GWs is only due to modifications in the MBH 
geometry.  This dephasing will not prevent in principle detection of GWs from EMRIs with LISA (from \emph{short} periods of data $\sim 3$ weeks,  
where radiation reaction effects can be neglected), 
but instead it will bias the estimation of parameters, leading to an uncontrolled systematic error.  

The study of radiation reaction effects in DCSMG~\cite{Sopuerta:2009iy} was carried out 
using the short-wave approximation (see, e.g.~\cite{Misner:1973cw}). It was found that 
to leading order the GW emission formulae are unchanged with respect to GR. 
That is, we can introduce an effective GW energy-momentum tensor that has exactly the same
form as in GR (the Isaacson tensor~\cite{Isaacson:1968gw}).  There are subdominant contributions
to the radiation reaction mechanism due to the presence of the DCSMG scalar field $\phi$.

By comparing waveforms computed in GR with waveforms computed in DCSMG 
(assuming the same orbital parameters: eccentricity, pericenter, and inclination), a 
a rough estimate~\cite{Sopuerta:2009iy} of the accuracy
to which DCSMG gravity could be constrained via a LISA observation was given. 
This estimate can be expressed as:
\begin{equation}
\xi^{1/4} \lesssim 10^{5}\, {\textrm{km}} \left(\frac{\delta}{10^{-6}}\right)^{1/4} 
\left(\frac{M^{}_{\bullet}}{M^{}_{\rm \tiny MW}}\right)\,,
\end{equation}
where $\delta$ is the accuracy to which $\xi$ can be measured, which depends on the integration time, 
the signal-to-noise ratio, the type of orbit considered and how much radiation-reaction affects the orbit.
Moreover, $M^{}_{\rm \tiny MW}$ is the mass of the presumable BH at Sgr A* in our Milky Way
galaxy, with a canonical value of $\sim 4.5 \cdot 10^{6}M^{}_{\odot}\,$.
Notice that IMRIs (with total masses in the range $10^3-10^{4} M^{}_{\odot}$) 
are favored over EMRIs. This result is to be compared 
with the binary pulsar constrained $\xi^{1/4} \lesssim 10^{4} \; {\textrm{km}}$~\cite{Yunes:2009hc}. 
These results imply that it may be possible to place strong constraints (up to two orders of 
magnitude more stringent than binary pulsar ones) with IMRI GW observations.
Moreover, a GW test can constrain the dynamical behavior of the theory in the neighbourhood 
of BHs, which is simply not possible with neutron star observations.

At present, there is work in progress that focuses on the inclusion of radiation reaction effects
and the use of better statistical tools to estimate the ability of LISA to constraint DCSMG.
A key point in this regard is that, to leading order, the GW emission in DCSMG is unchanged with
respect to GR, which can be used to simplify the analysis, allowing for GR-like 
expressions for the rate of change of constants of motion due to the GW emission.

\begin{acknowledgement}
CFS acknowledges support from the Ram\'on y Cajal Programme of the
Spanish Ministry of Education and Science (MEC) and by a Marie Curie
International Reintegration Grant (MIRG-CT-2007-205005/PHY) within the
7th European Community Framework Programme. 
Financial support from the contracts ESP2007-61712 (MEC), FIS2008-06078-C03-01 
and FIS2008-06078-C03-03 (MICINN) is gratefully acknowledged.
NY acknowledges support from NSF grant PHY 07-45779.
\end{acknowledgement}

\end{document}